\documentstyle[12pt,twoside]{article}

\catcode`\@=11
\def\newsymbol#1#2#3#4#5{\let\next@\relax%
 \ifnum#2=\@ne\else%
 \ifnum#2=\tw@\let\next@\msyfam@\fi\fi%
 \mathchardef#1="#3\next@#4#5}
\def\mathhexbox@#1#2#3{\relax%
 \ifmmode\mathpalette{}{\m@th\mnnathchar"#1#2#3}
 \else\leavevmode\hbox{$\m@th\mathchar"#1#2#3$}\fi}
\font\tenmsy=msbm10
\font\sevenmsy=msbm7
\font\fivemsy=msbm5
\newfam\msyfam
\textfont\msyfam=\tenmsy
\scriptfont\msyfam=\sevenmsy
\scriptscriptfont\msyfam=\fivemsy
\edef\msyfam@{\hexnumber@\msyfam}
\def\Bbb#1{\fam\msyfam\relax#1}
\catcode`\@=\active
\load{\footnotesize}{\sf}

\newtheorem{theorem}{Theorem}[section]
\newtheorem{proposition}[theorem]{Proposition}
\newtheorem{lemma}[theorem]{Lemma}
\newtheorem{corollary}[theorem]{Corollary}

\newtheorem{remark}[theorem]{Remark}

\newcommand{\eq}[1]{\begin{equation}\label{#1}}
\newcommand{\be}{\begin{eqnarray*}}
\newcommand{\ee}{\end{eqnarray*}}
\newcommand{\bee}{\begin{eqnarray}}
\newcommand{\eee}{\end{eqnarray}}
\newcommand{\en}{\end{equation}}
\newcommand{\proof}{{\noindent \it Proof:\ }}
\newcommand{\qed}{\hfill $\Box$\par\medskip}
\newcommand{\bi}{\begin{description}}
\newcommand{\bl}[1]{\begin{lemma}\label{#1}}
\newcommand{\el}{\end{lemma}}
\newcommand{\bt}[1]{\begin{theorem}\label{#1}}
\newcommand{\et}{\end{theorem}}
\newcommand{\bp}[1]{\begin{proposition}\label{#1}}
\newcommand{\ep}{\end{proposition}}
\newcommand{\bc}[1]{\begin{corollary}\label{#1}}
\newcommand{\ec}{\end{corollary}}

\newcommand{\BR}{{{\Bbb R}^3}}
\newcommand{\dr}{{\rm d} r}
\newcommand{\ei}{\end{description} }
\newcommand{\Y}{\int_{-1}^1\!\! {\rm d}X\!\! \int_\kappa^\La\!\! {\rm d}r}
\newcommand{\YY}{\int_{-1+(1/\La)}^0\! {\rm d}X \! \int_\kappa^\La \! {\rm d}r}
\newcommand{\YYY}{\int_0^{1-(1/\La)}\!  dy }
\newcommand{\YYYY}{\int_{15/16}^{1-(1/\La)}\! {\rm d}y }
\newcommand{\III}{\int_{-1}^1\!\! {\rm d}X\!\! \int_0^\La\!\! {\rm d}r}
\newcommand{\iint}{\int\!\!\!\int\!\! {\rm d^3} k_1{\rm d^3}k_2}
\newcommand{\II}{\int_\kappa^\La\!\! {\rm d}r}
\newcommand{\M}[1]{\left[#1\right]_\kappa^\La}
\newcommand{\N}{\M{\arctan\frac{r+\La X+1}{\sqrt\Delta}}}
\newcommand{\RN}{\M{\sqrt\Delta \frac{r+\La X+1}{\rho}}}
\newcommand{\nn}{\nonumber}
\newcommand{\NN}{\M{\arctan\frac{r-\La y+1}{\sqrt\Delta}}}
\newcommand{\RNN}{\M{\sqrt\Delta \frac{r-\La y+1}{\rho}}}
\newcommand{\QQ}{\frac{1}{(r_1^2+r_2^2+2r_1r_2X)/2+r_1+r_2}}
\newcommand{\rr}{\frac{1}{\rho_\La (r,X)}}
\newcommand{\rrr}{\rho}
\newcommand{\rrrr}{\int_\kappa ^\La \!\!\! {\rm d}r_1\!\! \int _\kappa^\La\!\!\!{\rm d}r_2}
\newcommand{\pr}{\pi r_1r_2}
\newcommand{\pa}{\pf\! \cdot\! A+A\!\cdot\! \pf}
\newcommand{\hi}{H_{\rm I}}
\newcommand{\hii}{H_{\rm II}}
\newcommand{\lz}{\left|}
\newcommand{\rz}{\right|}
\newcommand{\hz}{H_0\!\f}
\newcommand{\ej}{e_j(k_1)\cdot e_{j'}(k_2)}
\renewcommand{\v}{\varphi}
\newcommand{\vr}{\varphi^\r}
\newcommand{\r}{\rho}
\renewcommand{\t}{\Psi^\mu}
\renewcommand{\aa}{(A^+\!\cdot\! A^+)}
\newcommand{\aaa}{(A^-\!\cdot\! A^-)}
\newcommand{\ada}{(A^+\!\cdot\! A^-)}
\newcommand{\ggg}{(\gr(0),\gr(0))}
\newcommand{\no}{\|\la\|}
\newcommand{\s}{\sigma}
\newcommand{\C}[2]{\footnotesize {\lk\!\!\!\begin{array}{c}   #1 \\ #2
\end{array}\!\!\!\rk}}
\newcommand{\tr}{{\rm tr}}
\newcommand{\trr}{T_R(r)}
\newcommand{\RR}{{\Bbb R}}
\newcommand{\La}{\Lambda}
\newcommand{\liml}{\lim_{\Lambda\rightarrow\infty}}
\newcommand{\limn}{\lim_{n\rightarrow\infty}}
\newcommand{\limt}{\lim_{t\rightarrow\infty}}
\newcommand{\limT}{\lim_{T\rightarrow\infty}}
\newcommand{\limR}{\lim_{\La\rightarrow\infty}}
\newcommand{\kak}[1]{(\ref{#1})}
\newcommand{\LR}{{L^2(\BR)}}
\newcommand{\fff}{{{\cal F}}}
\newcommand{\fffk}{{{\cal F}}_\kappa}
\newcommand{\lec}{\lceil_{\fffk}}
\newcommand{\fffkk}{{{\cal F}}_{\kappa,0}}
\newcommand{\ffff}{{{\cal F}_{\rm fin}}}
\newcommand{\hhh}{{{\cal H}}}
\newcommand{\D}{\Delta}
\newcommand{\dk}{{\rm d}^3k}
\newcommand{\k}{\kappa}
\newcommand{\is}{\inf\sigma}
\newcommand{\f}{^{-1}}
\renewcommand{\P}[1]{\Phi_{#1}^\mu}
\newcommand{\lk}{\left(}
\newcommand{\rk}{\right)}
\newcommand{\lkk}{\left\{}
\newcommand{\rkk}{\right\}}
\newcommand{\lkkk}{\left[}
\newcommand{\rkkk}{\right]}
\newcommand{\wick}[1]{:\!\!{#1}\!\!:}
\newcommand{\XX}{\int_{-1}^1\!\!\!{\rm d}X}
\newcommand{\dx}{{\rm d}X}
\newcommand{\cx}{\cdots {\rm d}X}
\newcommand{\K}[1]{\frac{|\vp_{#1}|^2}{2\omega_{#1}}}
\newcommand{\KK}[1]{\frac{1}{E_{#1}}}
\newcommand{\KKK}{\frac{1}{E_{12}}}
\newcommand{\LL}[1]{\frac{1}{F_{#1}}}
\newcommand{\LLL}{\frac{1}{F_{12}}}
\newcommand{\qij}{(Q_1Q_2)}
\newcommand{\qijk}{Q_1Q_2}
\newcommand{\qi}{Q_1}
\newcommand{\qj}{Q_2}
\newcommand{\add}{a^{\ast}}
\newcommand{\ad}{a^\ast}
\newcommand{\ass}{a^\sharp}
\newcommand{\ov}[1]{\overline{#1}}
\newcommand{\hf}{H_{\rm f}}
\newcommand{\pf}{{P_{\rm f}}}
\newcommand{\si}[1]{\sigma(#1)}
\newcommand{\mmm}{\sum_{\mu=1}^3}
\newcommand{\gr}{\psi_{\rm g}}
\newcommand{\grk}{\psi_{{\rm g},\kappa}}
\newcommand{\grr}{\gr^\r}
\newcommand{\half}{\frac{1}{2}}
\newcommand{\han}{{1/2}}
\newcommand{\vp}{\hat{\varphi}}
\newcommand{\Av}{A_{\vp}}
\newcommand{\Hv}{H_{\vp}}
\newcommand{\Ev}{{E_{m,\La}}}
\newcommand{\Evv}{{E_{\vp}}}
\newcommand{\mass}{m_{\rm eff}}
\newcommand{\jj}{\sum_{j=1,2}}
\newcommand{\jjj}{\sum_{j,j'=1,2}}

\makeatletter
\@addtoreset{equation}{section}
\makeatother
\def\theequation{\arabic{section}.\arabic{equation}}

\begin{document}
\setlength{\baselineskip}{18pt}
\title{Mass renormalization in  nonrelativistic QED}
\author{Fumio Hiroshima \footnote{Department of Mathematics and Physics,
Setsunan University, 572-8508, Osaka, Japan. \hspace{5cm}
 email: hiroshima@mpg.setsunan.ac.jp}
and Herbert Spohn \footnote{Zentrum Mathematik and Physik 
Department, TU M\"unchen, D-80290,
M\"unchen, Germany. \hspace{3cm} email: spohn@ma.tum.de} }
\date{\today }
\maketitle
\begin{abstract}
In nonrelativistic QED the charge of an electron equals its bare
value, whereas the self-energy and the mass have to be
renormalized. In our contribution we study perturbative mass
renormalization, including second order in the fine structure
constant $\alpha$, in the case of a single, spinless electron. As
well known, if $m$ denotes the bare mass and $\mass$ the mass
computed from the theory, to order $\alpha$ one has
$$\frac{\mass}{m} =1+\frac{8\alpha}{3\pi}  \log(1+\half (\Lambda/m))+O(\alpha^2)$$
which suggests that $\mass/m=(\Lambda/m)^{8\alpha/3\pi}$ for small
$\alpha$. If correct, in order $\alpha^2$ the leading term should
be $\displaystyle \half ((8\alpha/3\pi)\log(\Lambda/m))^2$. To
check this point we expand $\mass/m$ to order $\alpha^2$. The
result is $\sqrt{\Lambda/m}$ as leading term, suggesting a more
complicated dependence of $m_{\mathrm{eff}}$ on $m$.
\end{abstract}
{\footnotesize
}

\section{Introduction}
Nonperturbative renormalization in relativistic QED remains as a
mathematical challenge. Thus it is of interest to study simplified
candidates, an obvious one being nonrelativistic QED. In this
theory, with comparable little effort, one can start from a
self-adjoint Hamiltonian operator and thus has a well-defined
mathematical framework. As an additional simplification, there is
no charge renormalization because of the absence of positrons.
Nevertheless, even in nonrelativistic QED, energy and mass
renormalization remain poorly understood. Our, admittedly modest,
contribution is to study mass renormalization including order
$\alpha^2$.

Let us first explain the basic Hamiltonian. We consider a single,
spinless free electron coupled to the quantized radiation field.
We will use relativistic units and employ immediately the total
momentum representation. Then the Hilbert space of  states is the
symmetric Fock space, $\fff$, over the one-particle space
$L^2(\RR^3\times \{1,2\})$, i.e.
$$\fff=\oplus_{n=0}^\infty\otimes_s^n L^2(\RR^3\times \{1,2\}).$$
The inner product in $\fff$ is denoted by $(\cdot, \cdot)$ and the
Fock vacuum by $\Omega$. On $\fff$ we introduce the Bose field
\eq{s1} a(f)=\jj\int f(k,j)^\ast a(k,j) dk,\ \ \ f\in L^2
(\RR^3\times \{1,2\}). \en Operators $a(f)$ and
$a(f)^\ast=a^\ast(f)$ are densely defined and satisfy the CCR
\begin{eqnarray*}
&&[a(f),a^\ast(g)]=(f,g)_{L^2(\RR^3\times \{1,2\})},\\
&&[a(f),a(g)]=0,\\
&&[a^\ast(f),a^\ast(g)]=0.
\end{eqnarray*}
The kinetic energy of the photon is given by \eq{s2}
\hf=\jj\int\omega(k)a^\ast(k,j)a(k,j) dk, \en which is the second
quantization of $\omega(k)=|k|$ considered as a multiplication
operator on $\LR$. Similarly the momentum of the photon field is
\eq{s3} \pf=\jj\int k a^\ast(k,j)a(k,j) dk. \en The coupling of
the electron to the Maxwell field is mediated through the
transverse vector potential $\Av$ defined by \eq{s4}
\Av=\frac{1}{\sqrt 2} (a(f)+a^\ast(f)), \en where \eq{s5}
f(k,j)=\frac{1}{\sqrt{\omega}}\vp(k) e(k,j) \en with $k/|k|,
e(k,1), e(k,2)$ forming a right-handed dreibein in $\BR$. $\vp$ is the form
factor which, as a minimal assumption, satisfies \eq{s166}
\int_\BR |\vp(k)|^2(\omega(k)^{-2}+\omega(k)) dk<\infty. \en Later
on, we will make more specific choice of $\vp$.

With these definitions the Hamiltonian under study is \eq{s6}
\Hv(p)=\half\wick{(p-\pf-e\Av)^2}+\hf,\ \ \ p\in\RR^3, \en where
$p$ is the total momentum,  $e$  the charge, to be definite $e\geq
0$, and $\wick{X}$ denotes the Wick order of $X$. 
$\Hv(p)$ with domain $D(\hf+\half\pf^2)=D(\hf)\cap D(\half \pf^2)$ is
self-adjoint for  $e$ and $p$ with $|e|<e_0$ and $|p|<p_0$ for some $e_0$ and $p_0$, 
provided \kak{s166} holds. The
energy-momentum relation is defined as the bottom of the spectrum
of $\Hv(p)$, \eq{s7} E_{\vp}(p)=\is (\Hv(p)). \en

In \kak{s6} the bare mass $m$ of the electron is still missing. It
appears in two places. Firstly the form factor depends on $m$. Let
us assume a sharp ultraviolet cutoff $\Lambda$. Then
\begin{eqnarray}
\label{s8}
&&\vp(k)=\vp_0(mck/\Lambda), \ \ \ \Lambda>0,\\
&&\vp_0(k)=\lkk
\begin{array}{ll}(2\pi)^{-3/2}& {\rm for}\  |k|\leq 1,\\
0& {\rm for}\  |k|>1,\end{array}
\right.\nn
\end{eqnarray}
with $1/mc$ the Compton wave length. Secondly energy is to be
measured in units of $mc^2$ and momentum in units of $mc$. We
henceforth set $c=1$ (and also $\hbar=1$). Thus the true
energy-momentum relation of the Pauli-Fierz Hamiltonian is \eq{s9}
\Ev(p)=mE_{\vp}(p/m),\ \ \ \vp \ {\rm of}\ \kak{s8}. \en Note that
equivalently $\Ev(p)$ is given through
$$\Ev(p)=\is(\frac{1}{2m}\!\wick{(p-\pf -eA_{\vp_0(\cdot/\Lambda)})^2}+\hf).$$

Removal of the ultraviolet cutoff $\Lambda$ through mass
renormalization means to find sequences \eq{s10}
\Lambda\rightarrow \infty,\ \ \  m\rightarrow 0 \en such that
$\Ev(p)-\Ev(0)$ has a nondegenerate limit. A convenient  criterion
for nondegeneracy is the curvature of $\Ev(p)$ at $p=0$, in other
words the inverse effective mass. Let us assume for a moment an
infrared cutoff
$$\vp(k)=0\ \ \ {\rm for}\ |k|<\kappa/m$$
with some $0<\kappa$. Then it is known \cite{hisp2} that, for
$|e|<e_\ast$, $|p|<p_\ast$ with suitable $e_\ast>0$ and
$p_\ast>0$, $\Hv(p)$ has a nondegenerate  ground state $\gr(p)$
separated by a gap from the continuum, i.e.
$$\Hv(p)\gr(p)=\Evv(p)\gr(p),\ \ \ \gr(p)\in\fff,$$
has a unique solution.
Let us set
\begin{equation}
\Ev(p)-\Ev(0)= \frac{1}{2\mass}p^2+ {\mathcal{O}}(|p|^3)
\end{equation}
for small  $p$.
Then, using second order perturbation theory in \kak{s9}, one obtains
\eq{s11}
\frac{m}{\mass}=1-\frac{2}{3}\sum_{\mu=1,2,3}
\frac{
(\gr(0), (\pf+e\Av)_\mu(\Hv(0)-\Evv(0))\f (\pf+e\Av)_\mu\gr(0))}{(\gr(0),\gr(0))}.
\en
We assume that this formula remains valid also when $\kappa=0$.

On the basis of \kak{s11}, mass renormalization can be discussed
more precisely. From \kak{s11} it trivially follows that $m/\mass$
depends only on the ratio $\La/m$. It is convenient to express
this dependence in the form \eq{s12} \frac{\mass}{m}=h(\La/m). \en
Clearly $h\geq 1$ and $h(0)=1$. Let us set \eq{s13}
\lambda=\frac{\La}{m} \en One expects that $h$ is increasing in
$\lambda$, because with increasing $\La$ more photons are bound to
the electron which makes $\mass$ larger.

Let us distinguish several cases. If $h$ has a finite limit as $\lambda\rightarrow \infty$, then
the mass renormalization is finite,
$$\mass=h(\infty)m.$$
Such kind of behavior occurs in the Nelson model \cite{ne3}.
Secondly let us consider the case that $h(\lambda)$ increases linearly for large $\lambda$.
We  set
$$h(\lambda)=1+b_0\lambda$$ with $b_0>0$. Then
\eq{s14} \mass=m+b_0\La. \en Hence mass renormalization is
additive. This behavior is found in the dipole approximation to
the Pauli-Fierz Hamiltonian, e.g. \cite{hisp1}, and in the
classical Abraham model \cite{sp}. If $\mass>0$ is imposed, then
as $\La\rightarrow \infty$ necessarily $m\rightarrow-\infty$. As
soon as $m<0$, $m\Hv(p/m)$ is not bounded from below. Therefore we
regard the theory as not renormalizable. Thus the case of interest
is when for large $\lambda$ \eq{s15} h(\lambda)\simeq
b_0\lambda^\gamma,\ \ \ b_0>0,\ \ \ 0<\gamma<1, \en which defines
the scaling exponent $\gamma$ and the amplitude $b_0$. $\gamma$
depends on $e$, as does $b_0$. Inserting \kak{s15} in \kak{s12},
one  obtains for sufficiently large $\lambda$, \eq{s16}
\frac{\mass}{m}\simeq b_0\lk\frac{\La}{m}\rk^\gamma. \en Thus the
choice \eq{s17} m=\La^{-\gamma/(1-\gamma)}b_1^{1/(1-\gamma)} \en
yields \eq{s18} \lim_{\Lambda\rightarrow
\infty}\mass(\La)=m^\ast=b_0b_1. \en Here $b_0$ is fixed by
$h(\lambda)$ and  $b_1$ is a parameter which can be adjusted to
yield the experimentally determined mass of the electron.

Of course, the difficulty with our discussion is that, while the
scaling function is well defined, at present we have no technique
to find out its behavior for large $\lambda$. For that reason we
turn to perturbative renormalization which, through the
interchange of the limits $\La\rightarrow \infty$  and
$e\rightarrow 0$, tries to guess the proper value of $\gamma$. The
details will be given in the following sections, but let us
summarize briefly our findings. The fine structure constant is
defined through \eq{s19} \alpha=\frac{e^2}{4\pi}. \en To first
order one finds \eq{s20}
h(\lambda)=1+\frac{8\alpha}{3\pi}\log(1+\half\lambda)+{\mathcal{O}}(\alpha^2),
\en which suggests \eq{s21} h(\lambda)\simeq
\lambda^{8\alpha/3\pi} \en for sufficiently large $\lambda$ and
therefore \eq{s22} \gamma=\frac{8\alpha}{3\pi},\ \ \ \ \alpha\ll
1. \en 
To have a control check, one assumes that to second order
\begin{eqnarray}
&&\hspace{-27pt}h(\lambda)\simeq \lambda^{(8/3\pi)\alpha+b\alpha^2}\nonumber\\
\label{s23}&& \simeq
1+\frac{8\alpha}{3\pi}\log\lambda+\half(\frac{8\alpha}{3\pi}
\log\lambda)^2+b\alpha^2\log\lambda+{\mathcal{O}}(\alpha^3)
\end{eqnarray}
for small $\alpha$. Therefore by expanding $\mass/m$ to order
$\alpha^2$, one should find a term $(\log \lambda)^2$ with an
already determined prefactor and a term propotional to
$\log\lambda$, together with lower order terms. As to explained,
this guess is not confirmed. {\it Instead} we prove that \eq{s24} 
h(\lambda)= 1+\frac{8\alpha}{3\pi}\log(1+\half\lambda)+c_0\alpha^2
\sqrt\lambda+{\mathcal{O}}(\alpha^3),\ \ \ c_0>0, \en 
for $|\alpha|$ small enough {\it depending on $\La$},  
which could
suggest $\displaystyle \gamma=\half$ independent of $e$.\medskip\\
\textit{Note added in proof}: Since the completion of this work F.H. and 
K. R. Ito \cite{hiit} extended the investigation 
to include the spin of the electron. 
The number of terms in the perturbation series up to the same order as studied here is then multiplied by a factor of 4. 
As a net result one finds that the leading divergence is proportional to $\Lambda^2$, 
rather than $\Lambda^{1/2}$. Because of the interaction with the quantized magnetic field the effective mass (at the order considered) is more strongly ultraviolet divergent when spin is included.\medskip

Some aspects of the effective mass  and its renormalization have
been studied before. Spohn \cite{sp5} investigates the effective
mass of the Nelson model \cite{ne3} from a functional integral
point of view. Lieb and Loss \cite{lilo, lilo2} study mass
renormalization and binding energies for various models of matter
coupled to the radiation field including the Pauli-Fierz model.
Hainzl \cite{ha1} and Hainzl and Seiringer \cite{hasi} compute the
leading order of the effective mass of the Pauli-Fierz Hamiltonian
with spin $\han$.

Our paper is organized in the following way. In Section 2 we
review under which conditions $\Evv(p)=\Evv(p,e)$ is jointly
analytic in $p$ and $e$. In Section 3 we set up the perturbation
theory for the effective mass and work out explicitely the terms
including $\alpha^2$. Their asymptotics with $\La\rightarrow
\infty$ is studied in Section 4.

\section{Ground state and its analytic properties}
Throughout this paper we assume that
$$
\vp(k)=\lkk
\begin{array}{ll}0& {\rm for}\ |k|<\kappa/m,\\
(2\pi)^{-3/2}& {\rm for}\  \kappa/m\leq |k|\leq \La/m,\\
0& {\rm for}\  |k|>\La/m.\end{array}
\right.
$$
For notational convenience, we shall use notations $H(p)$, $A$ and
$E(p)$ instead of $\Hv(p)$, $\Av$  and $\Evv(p)$, respectively.

Let $\fffk$ (resp. $\fffkk$)  
be the symmetric Fock space over $L^2(\RR^3_{\kappa/m}\times\{1,2\})$ (resp. 
$L^2(\RR^{3\perp}_{\kappa/m}\times\{1,2\})$), 
where 
$\RR^3_{\kappa/m}=\{k\in\BR||k|\geq \kappa/m\}$. 
Then it follows that 
\eq{j6}
\fff\cong \fffk\otimes\fffkk.
\en 
It is seen that $\fffk$ reduces $H(p)$ and,  under the identification \kak{j6}, 
\eq{j1}
H(p)\cong (H(p)\lceil_{\fffk})\otimes 1+1\otimes (\hf\lceil_{\fffkk}).
\en 
The bottom of the continuous
spectrum of $H(p)\lceil_{\fffk}$ is denoted by  
$E_c(p)$. Note that 
$\is(H(p)\lceil_{\fffk})=E(p)$. 
The following lemma can
be proven in the similar manner  as in \cite{fr2}.

\bl{3}{\rm \cite{fr2}}
There exists a constant  $p_\ast>0$ independent of $e$ with $|e|<e_0$ 
such that for $p\in\BR$ with $|p|<p_\ast$,
$$E_c(p)-E(p)>0.$$
In particular
there exists a  ground state $\grk(p)\in\fffk$
of $H(p)\lec$ for $p\in\BR$ provided   $|p|<p_\ast$.
\el
By Lemma \ref{3}, we see that 
$H(p)$ has the  ground state 
$$\gr(p)=\grk(p)\otimes\Omega_{\kappa,0}$$ 
for $p\in\BR$ provided   $|p|<p_\ast$, 
where $\Omega_{\kappa,0}$ denotes the vacuum of $\fffkk$.  
To have uniqueness, 
one proves that for any ground state $\gr(p)$,
one has $$(\gr(p), \Omega)>0$$
 provided $|p|<p_\ast$ and
$|e|<e_\ast$ with some $e\ast$. 

\bl{4} {\rm \cite{hisp2}} There exists a
constant $e_\ast>0$ such that
 for  $(p,e)\in\BR\times\RR$ with $|p|<p_\ast$ and
$|e|<e_\ast$,
the ground state of $H(p)$ is unique up to multiple constants.
\el

\begin{remark}
In the case  $\kappa=0$ and for sufficiently small $e$,
Chen \cite{ch} proves  the absence of a ground state of
$H(p)$ in $\fff$ for $p\not=0$
 and the existence of a ground state
of $H(0)$.
\end{remark}

We also need the analytic properties of $\gr(p)=\gr(p,e)$ and
$E(p)=E(p,e)$ with respect to $(p, e)\in \BR\times \RR$ in a neighborhood of 
$(0,0)\in\BR\times\RR$.

\bl{5}
There exists an open neighborhood ${\cal O}$ of  $(0,0)\in \BR\times\RR$
such that
$\gr(p,e)$  is strongly analytic
and $E(p,e)$  analytic on ${\cal O}$.
\el
\proof
Let $\grk(p)\in\fffk$ be the ground state of $H(p)\lceil_{\fffk}$. 
Since $\gr(p)=\grk(p)\otimes\Omega_{\kappa,0}$, it is enough to show that 
$\grk(p)$ is strongly analytic on ${\cal O}$. 
We split  $H(p)$ as
\eq{as}
H(p)=H_0(p)+ H_I(p),
\en
where
\be
& & H_0(p)=\half(p-\pf)^2 +\hf,\\
& & H_I(p)=-e(p-\pf)\cdot A+ e^2\frac{1}{2}\wick{A^2}.
\ee
Then we obtain that
\eq{pp1}
\|H_I(p)\Psi\|_{\fffk}\leq c_4\|H_0(p)\Psi\|_{\fffk}+c_5\|\Psi\|_{\fffk}
\en
for $\Psi\in D(H_0(p)\lceil_{\fffk})=D(\hf)\cap D(\pf^2)\cap\fffk$.
Then $H(p)\lec$ is an analytic family of type (A) for $e$ near
$e=0$ (see \cite[p.16] {rs4}).
Thus by \cite[Theorem XII.9]{rs4}, $H(p)\lec$ is an analytic family in the sense of Kato, which implies that
by \cite[Theorem XII.8]{rs4}, together with Lemmas \ref{3} and \ref{4},
$\grk(p,e)$ is strongly analytic
and $E(p,e)$ analytic for $e$ near $e=0$.
Alternatively we split  $H(p)$ as
$$H(p)=H'_0+p\cdot  H'_I+\half p^2,$$
where
$$ H'_0=\half\!\wick{(\pf+ e \Av)^2}+\hf,\ \ \  H'_I=-(\pf+ e \Av).$$
Then we have
\eq{pp2}
\|H_I' \Psi\|_{\fffk}\leq c_6\|H_0'\Psi\|_{\fffk}+c_7\|\Psi\|_{\fffk}
\en
with some constants $c_6$ and $c_7$ for $\Psi\in  D(\hf)\cap D(\pf^2)\cap \fffk$.
Thus
$H(p)\lec$ is an analytic family of type (A)
for $p\in\BR$ near $p=0$.
We can see that
$\grk(p,e)$ is strongly analytic and $E(p,e)$ analytic
for $p$ near $p=0$ in the similar manner as  for~$e$.
\qed

\section{Effective mass}
\subsection{Formulae}
In what follows  we assume that $(p,e)\in{\cal O}$. 
By the definition of $E(p,e)$, we have 
\eq{as2} \frac{m}{\mass}= \left. \frac{1}{3} \Delta_p E(p,e)
\right\lceil_{p=0}. \en 
Actually we can see in \cite{hisp2} that 
$H(p)$ is unitarily equivalent to $H(|p| n_z)$, where $n_z=(0,0,1)$. 
Thus $E(p,e)=\tilde E(|p|,e)=\is(H(|p| n_z))$ and 
$$\frac{m}{\mass}=\partial_{|p|}^2\tilde E(|p|,e)\lceil_{|p|=0}.$$ 
Moreover we see that 
$\tilde E(-|p|,e)=\tilde E(|p|,e)$. 
Then 
\eq{32} \left.
{\partial {p_\mu}} E(p,e)
\right\lceil_{p_\mu=0}=0,\ \ \ \mu=1,2,3. 
\en  
Since $E(p,e)$ also has the symmetry, $E(p,-e)=E(p,e)$,  
$E(p,e)$ is a function of $e^2$. 
In particular it follows that 
 \eq{odd} 
\left. \partial_e^{2m+1}
E(p,e)\right\lceil_{e=0}=0,\ \ \ m\geq 0. \en \bl{6} We have
$$
\frac{m}{\mass}=1-\frac{2}{3}\sum_{\mu=1,2,3}
\frac{(\gr(0), (\pf+eA)_\mu(H(0)-E(0))\f (\pf+eA)_\mu\gr(0))}{\ggg}.
$$
\el \proof 
Since
$$(H(p)\Psi, \gr(p))=E(p)(\Psi, \gr(p)),$$
for $\Psi\in D(H(p))$, 
taking a derivative with respect to $p_\mu$ onthe both sides above, 
we have
\eq{m4}
(H'_\mu(p)\Psi, \gr(p))+(H(p)\Psi, {\gr'}_\mu(p))={E'}_\mu(p)(\Psi, \gr(p))+E(p)(\Psi, {\gr'}_\mu(p))
\en
and
\begin{eqnarray}
&& (H''_\mu(p)\Psi, \gr(p))+2(H'_\mu(p)\Psi, {\gr'}_\mu(p))+(H(p)\Psi, {\gr''}_\mu(p))\nonumber\\
\label{33}
&& =E''_\mu(p)(\Psi, \gr(p))+2E'_\mu(p)(\Psi, {\gr'}_\mu(p))+E(p)(\Psi, {\gr''}_\mu(p)).
\end{eqnarray}
Here $E'_\mu(p)$ (resp. ${\gr'}_\mu(p)$) denotes the derivative (resp. strong derivative ) in $p_\mu$, 
and 
$H_\mu'(p)=(p-\pf-e\Av)_\mu$,  $H_\mu''(p)=1$. 
By \kak{32} it follows that
$E_\mu'(0)=0$,
and  by \kak{m4} with $p=0$,
\be
& &(\pf+e A)_\mu\gr(0)\in D((H(0)-E(0))\f),
\ee
and 
\be 
& & {\gr'}_\mu(0)=(H(0)-E(0))\f (\pf+e A)_\mu\gr(0). \ee
Therefore, using \kak{as2} and \kak{33}, we have 
\begin{eqnarray}
\frac{m}{\mass}&=&  \frac{1}{3}\sum_{\mu=1,2,3}
\frac{(\gr(0), E''(0)_\mu \gr(0))}{\ggg}\nonumber \\
&=& \label{j7}\frac{1}{3}\sum_{\mu=1,2,3}\lkk
1+\frac{(\gr(0), 2 H'_\mu(0){\gr'}_\mu(0))}{\ggg}\rkk \\
&=& 1 - \frac{2}{3} \sum_{\mu=1,2,3} \frac{ \lk (\pf+e A)_\mu
\gr(0), (H(0)-E(0))\f (\pf+e A)_\mu \gr(0)\rk}{\ggg}\,. \nonumber 
\end{eqnarray} 
Thus the lemma follows. \qed From this lemma we obtain the
following corollary. \bc{sp1} Let $|e|<e_\ast$.  Then $ \mass\geq
m$. \ec

\subsection{Perturbative expansions}
Let
$$
\gr(0)=\sum_{n=0}^\infty \frac{e^n }{n!} \v_n,\ \ \
E(0)=
\sum_{n=0}^\infty \frac{e^{2n}}{(2n)!} E_{2n}.
$$
We want to get the explicit form of $\v_n$.
Let
\be
&& 
\ffff=\{\{\Psi^{(n)}\}_{n=0}^\infty\in\fff|
\Psi^{(m)}=0 \mbox{ for } m\geq \ell\mbox { with some }
\ell\},\\
&&
\fff_0=\lkk
\left. \Psi \in\ffff \right | {\rm (i)}\  \Psi^{(0)}=0,
{\rm (ii)}\
{\rm supp}_{k\in\RR^{3n}} \Psi^{(n)}(k, j)\not\ni \{0\}, n\geq1, j\in\{1,2\}^n\rkk.
\ee

\bl{111}
We see that
$\fff_0\subset D(H_0\f).$
\el
\proof
Let $\Psi=\{\Psi^{(n)}\}_{n=0}^\infty\in\fff_0$.
Since
\be
&&   (\hz \Psi)^{(n)}(k_1,...,k_n, j_1,...,j_n)\\
&& = 
\lkkk\half(k_1+\cdots+k_n)^2+\sum_{i=1}^n\omega(k_i)\rkkk\f
\Psi^{(n)}(k_1,...,k_n,j_1,...,j_n), 
\ee and 
${\rm supp}_{(k_1,...,k_n)\in \RR^{3n}}
 \Psi^{(n)}(k_1,...,k_n,j_1,...,j_n)\not\ni
\{(0,...,0)\}$,
we obtain that
$$\|\hz \Psi\|_\fff^2=\sum_{n=1}^\infty \|(\hz \Psi)^{(n)}\|_{\fff^{(n)}}^2
<\infty$$ and the lemma follows. \qed
We define $A^-$ and $A^+$ by
\be
 A^-=\frac{1}{\sqrt2} a(f),\ \ \ A^+=\frac{1}{\sqrt2} \add(f).
\ee
Then
$A=A^++A^-$.
Moreover $A_\mu^-$ and $A_\mu^+$ are defined by $A^-$ and $A^+$ with
$e(k,j)$ replaced by $e^\mu(k,j)$.
We split
$H(0)$ as
$$H(0)=H_0 +e H_1+\frac{e^2}{2}{H_2}, $$
where
\be
& & H_0=\half \pf^2+\hf,\\
& & H_1=\half(\pf\cdot A+A\cdot \pf)=\pf \cdot A=A\cdot \pf,\\
& & H_2=\wick{A^2}=A^{+}\cdot A^++A^-\cdot A^-+2A^+\cdot A^-.
\ee
\bl{9}
We have $E_0=E_2=0$,  and there exists a ground state 
$\displaystyle \gr(0)=\sum_{n=0}^\infty \frac{e^n }{n!} \v_n$ 
such that 
\eq{j8}
\v_0=\Omega,\ \ \ \v_1=0,\ \ \
\v_2=-\hz H_2\Omega,\ \ \
\v_3=3\hz H_1 \hz H_2\Omega .
\en 
In particular
$\v_2\in \fff^{(2)}$ and $\v_3\in\fff^{(1)}\oplus \fff^{(3)}$.
\el
\proof
It is obvious that
$E_0=0$. 
Let 
$\displaystyle \gr(0)=\sum_{n=0}^\infty \frac{e^n }{n!} \v_n$ 
be an arbitrary strongly analytic ground state of $H(0)$ with 
$(\v_0, \Omega)\not=0$.  
Let $\displaystyle \rho(e)=\sum_{n=0}^\infty \frac{e^n }{n!} \r_n$  
be an analytic function on $e$.  
Then $\rho\gr(0)$ is also a strongly analytic  ground state of $H(0)$ and 
\be\rho\gr(0)
&=& \underbrace{\r_1\v_0}_{=\vr_0}+e\underbrace{
(\r_0\v _1+\r_1\v_0)}_{=\vr_1}+e^2 \frac{1}{2!}\underbrace{
(\r_0\v_2+2\r_1\v_1+\r_2\v_0)}_{=\vr_2}\\
&& +e^3\frac{1}{3!}\underbrace{
(\r_0\v_3+3\r_1\v_2+3\r_2\v_1+\r_3\v_0)}_{=\vr_3}+{\mathcal O}(e^4).
\ee
Set
\be
&& \r_0=1/(\v_0,\Omega),\ \ \  \r_1=-\r_0(\v_1,\Omega)/(\v_0,\Omega), \\
&& 
\r_2=-(\r_0(\v_2,\Omega)+2\r_1(\v_1,\Omega))/(\v_0,\Omega),\\
&& \r_3=-(\r_0(\v_3,\Omega)+3\r_1(\v_2,\Omega)+3\r_2(\v_1,\Omega))/(\v_0,\Omega).
\ee
Then 
$\displaystyle \grr=\sum_{n=0}^\infty \frac{e^n }{n!} \vr_n$ 
satisfies that 
\eq{j3}
(\vr_n, \Omega)=\delta_{0,n},\ \ \ n=0, 1,2,3.
\en 
We reset $\grr$ (resp. $\vr_n$)  with \kak{j3} as $\gr(0)$ (resp. $\v_n$). 
Let us write 
$H(0)$, $E(0)$ and $\gr(0)$  as $H$, $E$ and $\gr$, respectively.
Take derivative in $e$ on the both sides of 
$(H\Psi, \gr)=E(\Psi, \gr)$, $\Psi\in D(H)$. 
Then we have 
\begin{eqnarray}
&&\hspace{-1.3cm} \label{11}
(H'\Psi, \gr)+(H\Psi, \gr')=E'(\Psi, \gr)+E(\Psi, \gr'),\\
&&\hspace{-1.3cm}\label{12}
(H''\Psi, \gr)+2(H'\Psi, \gr')+(H\Psi, \gr'')=E''(\Psi, \gr)+2E'(\Psi, \gr')+E(\Psi, \gr''),\\
&&
\hspace{-1.3cm}
3(H''\Psi, \gr')+3(H'\Psi, \gr'')+(H\Psi, \gr^{'''})\nonumber \\ 
&&\hspace{2.5cm}\label{13}=
E^{'''}(\Psi, \gr)+3E''(\Psi, \gr')+3E'(\Psi, \gr'')+E(\Psi, \gr^{'''}), 
\end{eqnarray}
where $E'$ (resp. $\gr'$) denotes the derivative (resp. strong derivative) in $e$, 
and 
$H'=\pf(\pf+e A)$ and $H''=\pf\cdot A$. 
Put $\Psi=\Omega$ and $e=0$ in \kak{12}. 
Then
\eq{39}
0=E_2(\Omega,\Omega), 
\en
which shows 
that $E_2=0$.
From \kak{11} with $e=0$ it follows that
$$H_1\Omega+H_0\v_1=0,$$
from which it holds  that
$ H_0\v_1=0$.
Hence 
$\v_1= b\Omega$
with some constant $b$.
By  \kak{j3} we have, however, 
$b=0$. Then 
$\v_1=0$ follows.
By \kak{12} with $e=0$,
we have
$$H_2\Omega+H_0\v_2=0.$$
Since $H_2\Omega\in\fff_0$,
we see that by Lemma \ref{111},
$H_2\Omega\in D(H_0\f)$.
Thus we have
$\v_2=-\hz H_2\Omega+c\Omega$ with some constant $c$.
Since $(-\hz H_2\Omega, \Omega)=0$, 
it follows  that $c=0$ from \kak{j3}. 
From \kak{13} 
it follows that
 in  $e=0$,
$$3H_1\v_2+H_0\v_3=0.$$
Since
$H_1\v_2=-H_1\hz H_2\Omega\in\fff_0$, Lemma \ref{111} ensures that
$H_1\v_2\in D(\hz)$.
Hence
$\v_3=-3 \hz H_1\v_2+d\Omega=
3\hz H_1 \hz H_2\Omega+d\Omega$ with some constant $d$.
Since $(3\hz H_1 \hz H_2\Omega, \Omega)=0$, 
it follows  that $d=0$  from  \kak{j3}. 
Then the lemma is proven.
\qed

In the similar manner as Lemma \ref{9}, we can prove the following
proposition. 
\bp{02} There exists a ground state 
$\displaystyle \gr(0)=\sum_{n=0}^\infty \frac{e^n }{n!} \v_n$ 
such that 
\be 
& & \hspace{-0.5cm}
\v_{2m}=\hz\lkk -\sum_{j=1,2}\C{2m}{j}H_1\v_{2m-j}
+\sum_{j=2}^m \C{2m}{2j}E_{2j}\v_{2m-2j} \rkk,\ \ \ m\geq 2, \\
&& \hspace{-0.5cm}
\v_{2m+1}=\hz\lkk
-\sum_{j=1,2} \C{2m+1}{j}H_j\v_{2m+1-j}
+\sum_{j=2}^{m-1}\C{2m+1}{2j}E_{2j} \v_{2m-2j+1}
\rkk, m\geq 2,
\ee
with 
$\v_{2m}\in \fff^{(2)}\oplus \fff^{(4)}\oplus \cdots \oplus
\fff^{(2m)}$ and 
$\v_{2m+1}\in \fff^{(1)}\oplus \fff^{(3)}\oplus \cdots \oplus \fff^{(2m+1)}$, 
and $E_{2m}$ is given by 
$$E_{2m}=\C {2m}{2} ( \Omega, H_2\v_{2m-2}),\ \ \ m\geq 2.$$
\ep

\subsection{Effective mass up to order $e^4$}
In this subsection we expand $m/\mass$ up to order $e^4$.
\bl{mainlemma}
We have
\bee
\frac{m}{\mass}&=&
1-
e^2   \frac{2}{3}  \sum_{\mu=1,2,3}
\lk\Omega, {A}_\mu \hz {A}_\mu
\Omega\rk \nn\\
& &
-
e^4 \frac{2}{3}   \sum_{\mu=1,2,3}
\lkk
2\lk \t_3,\hz\t_1\rk +
\lk \t_2, \hz \t_2\rk -
2\lk \t_2, \hz H_1 \hz \t_1\rk\right.\nn\\
\label{c4}
& & \left.
\hspace{0.7cm}
-\half \lk \t_1, \hz  H_2 \hz \t_1\rk+
\lk \t_1,\hz H_1 \hz  H_1 \hz \t_1\rk\rkk+ {\mathcal{O}}(e^6),\nn\\
&& \label{313}
\eee
where
\be
\t_1&=& {A}_\mu\Omega,\\
\t_2&=&-\half \pf_\mu\hz \aa\Omega,\\
\t_3&=&\half
\lkk
-{A}_\mu\hz\aa\Omega+\half \pf_\mu \hz(\pa)\hz\aa\Omega\rkk .
\ee
\el
\proof
Since by \kak{j7}, 
\begin{eqnarray}
\label{q0}
\frac{m}{\mass}
=
1-\frac{2}{3}\sum_{\mu=1,2,3}
\frac{((\pf+eA)_\mu \gr(0), {\gr'}_\mu(0))}{\ggg}, 
\end{eqnarray} 
where ${\gr'}_\mu(0)={\rm s}-\partial_{p_\mu}\gr(p)\lceil_{p=0}$, 
we expand ${\gr'}_\mu(0)$ and $\gr(0)$ in $e$. 
Assume that 
$\displaystyle \gr(0)=\sum_{n=0}^\infty\frac{e^n}{n!}\v_n$  satisfies \kak{j8}, 
i.e., $\v_0=\Omega$, $\v_1=0$, 
$\v_2=-\hz H_2\Omega$ and 
$\v_3=3\hz H_1 \hz H_2\Omega$. 
We have 
$$ (\pf+e  A)_\mu
\gr(0)=e  {A}_\mu \Omega+e^2(\half \pf_\mu \v_2)+
e^3(\half  {A}_\mu \v_2+\frac{1}{6}  \pf_\mu \v_3)+
{\mathcal{O}} (e^4)$$ \eq{317} =e \t_1+e^2\t_2+e^3 \t_3+
{ \mathcal{O}} (e^4). \en 
Note that by Proposition \ref{02},
$$\v_0\in\fff^{(0)},
\v_2\in\fff^{(2)}, \v_3\in\fff^{(3)}\oplus \fff^{(1)},
\v_4\in\fff^{(4)}\oplus \fff^{(2)}.$$ In particular \eq{r1}
\frac{1}{(\gr, \gr)}= 1-e^4(\half\v_2,\half\v_2)-
e^4(\Omega, \frac{1}{24}\v_4)+ {\mathcal{O}}
(e^6)=1-e^4\frac{1}{4}(\v_2,\v_2) + { \mathcal{O}}(e^6).
\en 
Let 
\eq{3171}
\displaystyle {\gr'}_\mu(0)=\sum_{n=0}^\infty \frac{e^n}{n!}\P n.
\en
Since 
\eq{N}
((H(0)-E(0))\Psi, {\gr'}_\mu(0))=((\pf+eA)_\mu\Psi, \gr(0)),\ \ \ \Psi\in D(H(0)),
\en 
putting $e=0$ on the both sides of \kak{N},  
we have 
$$H_0\P 0=0.$$ 
Then 
\eq{j10}
\P 0=b_0\Omega
\en  with some constant $b_0$.  
From taking derivative  of the both sides of   \kak{N} 
at $e=0$, 
we see that  by \kak{j8}
\be
&& H_0\P 1=A_\mu\Omega,\\
&& H_2 \P 0+2 H_1 \P 1+H_0\P 2=\pf_\mu\v_ 2,\\
&& 3H_2 \P 1+3 H_1 \P 2+H_0\P 3=3A_\mu\v_ 2+\pf_\mu\v_ 3.
\ee
From them it follows that 
\begin{eqnarray} 
\label{q1} 
\P 1 &=& \hz \t_1+b_1 \Omega, \\
\P 2 &=& \hz(2\t_2-2H_1\P 1-H_2\P 0)+b_2\Omega,\nn \\
&=& \label{q2} 
2 \hz(\t_2-H_1\hz\t_1)+(-b_0\hz H_2\Omega+b_2\Omega)\\
\P 3
&=& \hz( 6\t_3-3H_1\P 2-3H_2\P 1)+b_3\Omega,\nn \\
&=& 6 \hz(\t_3-H_1\hz\t_2+(H_1\hz H_1\hz-\half H_2\hz)\t_1)\nn \\
 & &\label{q3}  +(3b_0\hz H_1\hz H_2\Omega-3b_1\hz H_2\Omega+b_3\Omega),
\end{eqnarray} 
where $b_1,b_2,b_3$ are some constants. Here we used that $H_1\Omega=0$. 
By \kak{q0}, \kak{317}, \kak{r1} and \kak{3171} we have 
\begin{eqnarray}
\frac{m}{\mass}
&=&1-\frac{2}{3}\mmm 
\lkk 
e^2\lk (\t_1, \P 1)+(\t_2,\P 0)\rk\right.\nonumber \\
&+& 
\label{q10}
\left. e^4\lk \frac{1}{6}(\t_1, \P 3)+\frac{1}{2}(\t_2, \P 2)+
(\t _3, \P 1)\rk\rkk+{\mathcal {O}}(e^6).
\end{eqnarray} 
Substitute \kak{q1}-\kak{q3} 
into \kak{q10}. 
No contribution of constants $b_0,...,b_3$ exists, i.e., 
we can directly see that 
$$ 
e^2 \lkk b_1 (\t_1, \Omega)+
b_0(\t_2,\Omega)\rkk=0.$$
and 
\be
&& 
e^4\lkk 
\frac{1}{6}
b_3(\t_1, \Omega)+
\frac{1}{6} 
b_1(\t_1, -3\hz H_2\Omega)
+
\frac{1}{6} 
b_0(\t_1, 3\hz H_1\hz H_2\Omega)\right.\\
&& 
\hspace{4cm}\left. 
+\half 
b_2 (\t_2,\Omega)+
\half b_0 (\t_2, -\hz H_2\Omega)
+b_1(\t_3,\Omega)\rkk =0.
\ee
Then the lemma follows. 
\qed

\begin{remark}
By Lemma \ref{6}
we have seen that
\eq{314}
\frac{m}{\mass}=1-\frac{2}{3}
\sum_{\mu=1,2,3}
\frac{
\lk (\pf+e  A)_\mu \gr(0), (H(0)-E(0))\f
(\pf+e A)_\mu \gr(0)\rk}{\ggg}.
\en
We  ``informally'' expand 
$(H(0)-E(0))\f $ as 
\begin{eqnarray}
&&\hspace{-1cm}
 (H(0)-E(0))\f
=\sum_{n=0}^\infty\lk-\hz\sum_{l=1}^\infty \frac{e^l}{l!}H_l\rk^n\hz\nonumber \\
&&\hspace{-1cm}\label{j4}= \sum_{n=0}^\infty (-1)^n\sum_{k=n}^\infty e^k
\sum_{\stackrel{l_1,...,l_n=1}{l_1+\cdots+l_n=k}}^k
\frac{1}{l_1!\cdots l_n!}
\hz H_{l_1}\hz H_{l_2}\cdots \hz H_{l_n}\hz.
\end{eqnarray}
Here we set
$H_j=\lkk\begin{array}{ll}H_j, & j=1,2,\\
-E_j, & j\geq 3.\end{array}\right.$ 
Then we have 
\begin{eqnarray}
 (H(0)-E(0))\f
&=& \hz-e
\hz H_1 \hz\nonumber \\
\label{315} & &  \hspace{-1.5cm}+e^2\lk -\half  \hz  H_2\hz+\hz
H_1 \hz H_1\hz\rk + {\mathcal{O}}(e^3).
\end{eqnarray}
Substitute \kak{315} into 
\kak{314}. Then the result coincides with \kak{313}. 
\end{remark}

\subsection{Explicit expressions}
For each $k\in\BR$ let us define  the projection $Q(k)$  on $\BR$ by
$$Q(k)=\jj |e_j(k)\rangle\langle e_j(k)|.$$
We also set
$$m=1,$$\\
since it can easily be reintroduced at the end of the computation.
We set
$$\vp_j=\vp(k_j),\ \ \omega_j=\omega(k_j),\ \ Q(k_j)=Q_j, \ \ \ j=1,2.$$
Let
\be
\KK j&=&\frac{1}{|k_j|^2/2+\omega_j},\ \ \ j=1,2,\\
\KKK&=& \frac{1} {|k_1+k_2|^2/2+\omega_1+\omega_2},\ \ \ k_1,k_2\in\BR,\\
\LL j&=&\frac{1}{r_j^2/2+r_j},\ \ \ j=1,2,\\
\LLL &=&\frac{1}{(r_1^2+2r_1r_2 X + r_2^2)/2+r_1+r_2},\ \ \ r_1,r_2\geq 0, \ \ -1\leq X \leq 1.
\ee
\bl{77}
We have
$$\frac{m}{\mass}=1-\alpha  a_1(\Lambda,\kappa) -\alpha^2 a_2(\Lambda,\kappa)+
{\mathcal{O}}(\alpha^3),$$ where \eq{a1}
a_1(\Lambda,\kappa)=\frac{8}{3\pi} \log\lk\frac{\La+2}{\kappa
+2}\rk \en and
\begin{eqnarray}
 a_2(\Lambda,\kappa)
&=&
(4\pi)^2 \frac{2}{3}
\iint \K 1 \K 2 \times \nn\\
& & \times \lkk
-\lk\KK 1+\KK 2\rk\KKK(1+s^2)+
\lk \KKK\rk^3 \frac{|k_1+k_2|^2}{2}(1+s^2)\right.\nn
\\
& & \left.+
\lk\KK 1+\KK 2\rk\lk\KKK\rk^2(k_1\cdot k_2)(-1+s^2)-\KK 1\KK 2(1+s^2)\right.\nn\\
& & \left. +
\lk \frac{|k_1|^2}{E_1^2}+\frac{|k_2|^2}{E_2^2}\rk\KKK (1-s^2)+\KK 1\KK 2 \KKK(k_1\cdot k_2)(-1+s^2)
\rkk,\nn \\
\label{79}&&
\end{eqnarray}
where $s=(\hat{k}_1, \hat{k}_2).$ Changing variables to polar
coordinates we also have
\begin{eqnarray}
 a_2(\Lambda,\kappa)
&=&
\frac{(4\pi)^2}{(2\pi)^6}
\frac{2}{3}
\XX \int_{\kappa}^{\Lambda} {\rm d}r_1\int_{\kappa}^{\Lambda}{\rm d}r_2 \pr \times \nn\\
& & \times \lkk
-\lk\LL 1+\LL 2\rk\LLL(1+X^2)+
\lk \LLL\rk^3 \frac{r_1^2+2r_1r_2 X+r_2^2}{2}(1+X^2)\right.
\nn\\
& & \left.+
\lk\LL 1+\LL 2\rk\lk\LLL\rk^2  r_1 r_2 X(-1+X^2)
-\LL 1\LL 2(1+X^2)\right.\nn\\
& & \label{80}
\left. +
\lk \frac{r_1^2}{F_1^2}
+\frac{r_2^2}{F_2^2}\rk\LLL (1-X^2)
+\LL 1\LL 2 \LLL r_1 r_2 X(-1+X^2)
\rkk.
\end{eqnarray}
\el
\proof
Note
that
\be a_1(\Lambda,\kappa)&=&
\frac{2}{3}(\sqrt{4\pi})^2
(A^+_\mu\Omega, \hz A^+_\mu\Omega)\\
 &=& \frac{2}{3} (\sqrt{4\pi})^2  2 \int\frac{\vp(k)^2}{2\omega(k)}\frac{1}{|k|^2/2+|k|} \dk\\
& =&  \frac{2}{3} (\sqrt{4\pi})^2 \frac{1}{(2\pi)^3} 4
\pi \int_{\kappa}^{\Lambda}\frac{1}{r/2+ 1} \dr\\
&=& \frac{8}{3\pi} \log\lk\frac{\La+2}{\kappa+2}\rk. \ee Thus
\kak{a1} follows. To check  $a_2(\Lambda, \kappa)$ we exactly
compute  the   five  terms on the right-hand side of \kak{c4}
separately.\medskip\\
{\bf (1)} We have
\bee
2\lk \t_3, \hz \t_1\rk
&=&
\lk \Omega, -\aaa \hz {A}_\mu \hz A^+_\mu  \Omega\rk\nn\\
& & +\half \lk \Omega, \aaa \hz (\pa)\hz \pf_\mu \hz
A^+_\mu \Omega\rk\nn.\\
& & \label{41}
\eee
Since
$\pf_\mu \hz {A}_\mu\Omega =\hz {A}_\mu \pf_\mu\Omega=0$,
the second term of the righ-hand side of \kak{41} vanishes, we have
\bee
2\lk \t_3,\hz \t_1\rk
&=&
-
\lk \Omega, \aaa\hz A^+_\mu \hz A^+_\mu\Omega\rk\nn\\
&=& -\iint  \K 1 \K 2 \KKK (\KK 1+\KK 2)  \tr \qij.\nn\\
& & \label{E1}
\eee
{\bf (2)} We have
\bee\lk \t_2, \hz \t_2\rk
&=&
\lk\frac{1}{2}\rk^2\lk \pf_\mu\hz \aa\Omega, \hz\pf_\mu \hz\aa\Omega\rk\nn\\
&=&
\lk\frac{1}{2}\rk^2 \lk \Omega, \aaa \lk \hz\rk^3 (\pf\cdot \pf) \aa\Omega\rk\nn\\
&=& \lk\frac{1}{2}\rk^2 \iint  \K 1 \K 2 \lk\KKK\rk^3 |k_1+k_2|^2
2 \tr \qij. \nn\\
&& \label{E2}
\eee
{\bf (3)} We have
\bee
& & -2 \lk \t_2, \hz H_1 \hz \t_1\rk\nn\\
& &=
\half \lk \pf_\mu\hz \aa\Omega, \hz(\pa)\hz A^+_\mu \Omega\rk\nn\\
& &=
\sum_{\nu=1,2,3}
\lk \Omega, \aaa\hz \pf_\mu \hz \pf_\nu A^+ _\nu  \hz
A^+_\mu \Omega\rk\nn\\
& &\label{E3} = \iint  \K 1 \K 2 \lk \KKK\rk^2 (\KK 1+\KK 2)
(k_2, \qijk k_1). \eee
{\bf (4)} We have
\bee
& & -\half \lk \t_1, \hz H_2 \hz \t_1\rk\nn\\
& & =
-\half \lk A^+_\mu \Omega, \hz (\aa+2\ada  +\aaa)\hz A^+_\mu
\Omega\rk \nn\\
& & =-\lk \Omega, A^-_\mu\hz \ada  \hz A^+_\mu  \Omega\rk \nn\\
\label{E4} & & =-\iint  \K 1 \K 2 \KK 1 \KK 2 \tr \qij . \eee
{\bf (5)} We have
\bee
& & \lk \t_1, \hz H_1 \hz H_1 \hz \t_1\rk\nn\\
& & = \lk\half \rk^2
\lk
A^+_\mu  \Omega, \hz (\pa) \hz (\pa) \hz A^+_\mu \Omega\rk \nn\\
& & =
\lk A^+_\mu\Omega, \hz (\pf\!\cdot \! A) \hz (\pf\! \cdot\! A) \hz
A^+_\mu\Omega\rk \nn\\
& & =\sum_{\nu,\kappa=1,2,3}
\lk A^+_\mu  \Omega, \hz \pf_\nu A^+_\nu \hz \pf_\kappa A^-_\kappa\hz A^+_\mu
\Omega\rk\nn\\
\label
{t2}
& & \hspace{2cm}
+\sum_{\nu,\kappa=1,2,3}
\lk A^+_\mu \Omega, \hz \pf_\nu A^-_\nu \hz \pf_\kappa A^+_\kappa\hz A^+_\mu
\Omega\rk.
\eee
Since
$$\pf_\kappa A_\kappa^-\hz A_\mu^+\Omega=0,$$
the  first term on the last line in \kak{t2} vanishes. Then we
have \bee
& & \lk \t_1, \hz H_1 \hz H_1 \hz \t_1\rk\nn\\
& & =\sum_{\nu,\kappa=1,2,3}
\lk  \Omega,  A^-_\mu\hz \pf_\nu A^-_\nu \hz \pf_\kappa A^+_\kappa\hz A^+_\mu
\Omega\rk\nn\\
& & = \iint  \K 1 \K 2 \KKK \lkk
\lk \KK 2\rk ^2 2 (k_2, \qi k_2) +\KK 1\KK 2 (k_2, \qijk  k_1)\rkk\nn\\
& & = \iint  \K 1 \K 2 \KKK \lkk
\lk \KK 1\rk ^2  (k_1, \qj k_1)+\lk \KK 2\rk ^2  (k_2, \qi k_2)\rkk\nn\\
\label{E5} & & \hspace{2cm}+ \iint  \K 1 \K 2 \KKK \KK 1\KK 2
(k_2, \qijk k_1). \eee  \kak{79}  follows from Lemma
\ref{mainlemma}, \kak{E1}, \kak{E2}, \kak{E3}, \kak{E4}, \kak{E5}
and the facts \be
\tr [Q(k_1)Q(k_2)]&=&\jjj(e_j(k_1)e_{j'}(k_2))^2=1+(\hat k_1, \hat k_2)^2,\\
(k_1, Q(k_2) Q(k_1) k_2)&=&
( k_1,  k_2)((\hat k_1, \hat k_2)^2-1),\\
(k_1, Q(k_2) k_1)&=&|k_1|^2(1-(\hat k_1, \hat k_2)^2).
\ee
Thus the proof is complete.
\qed

\section{Main theorem}
By \kak{80} we can see that
\eq{imamiya}
a_2(\La ,\kappa)=\frac{(4\pi)^2}{(2\pi)^6}\frac{2}{3}\sum_{j=1}^6 b_j(\La ),
\en
where
\be
& & b_1(\La )=-\XX (1+X^2)\rrrr \pr \lk\LL 1+\LL 2\rk\LLL,\\
& & b_2(\La )=\XX  (1+X^2)\rrrr \pr
\lk \LLL\rk^3 \frac{r_1^2+2r_1r_2 X+r_2^2}{2},\\
& & b_3(\La)=\XX  X(-1+X^2) \rrrr \pi r_1^2r_2^2
\lk\LL 1+\LL 2\rk\lk\LLL\rk^2 ,\\
& & b_4(\La)= -\XX (1+X^2)\rrrr \pr  \LL 1\LL 2,\\
& & b_5(\La)= \XX  (1-X^2) \rrrr \pr\lk \frac{r_1^2}{F_1^2}
+\frac{r_2^2}{F_2^2}\rk\LLL, \\
& & b_6(\La)=\XX   X(-1+X^2) \rrrr \pi r_1^2r_2^2   \LL 1\LL 2
\LLL. \ee Our main theorem  is stated as follows. \bt{main} There
exist strictly positive constants $c_1$ and $c_2$ such that
$$c_1\leq \liml \frac{a_2(\Lambda, \kappa)}{\sqrt{\Lambda}}\leq c_2.$$
\et
To prove Theorem \ref{main}  we estimate the lower and upper bounds of
$a_2(\La,\kappa)/\sqrt\La$ as $\La\rightarrow \infty$ in what follows.

Let
$\rho_\La(\cdot,\cdot):[0,\infty)\times [-1,1]\rightarrow \RR$ be defined by
$$\rho_\La=\rho_\La(r, X)=r^2+  2\La rX +\La^2 +2r+2\La=(r+\La X+1)^2+\Delta,$$
where
\eq{D}
\Delta=\La^2(1-X^2)+2\La(1-X)-1.\en
\bl{hir1}
There exist constants $C_1,C_2,C_3$ and $C_4$ such that
for sufficiently large $\La>0$,
\be
& &  (1)\ \III  \rr \leq C_1\frac{1}{\La },\\
& &
(2)\  \III \lk \rr\rk ^2 \leq C_2 \frac{1}{\La ^{5/2}},\\
& & (3)\  \III \rr\frac{1}{r+2}\leq C_3 \frac{\log \La }{\La ^2}, \\
& &
(4)\ \III  \lk \rr\rk^2(1-X^2) \leq C_4 \frac{1}{\La ^3}.
\ee
\el
\proof
See Appendix A.
\qed

\subsection{Upper bounds}
\bl{H3}
There exists a constant $C_{\rm max}$ such that
$$\limR \left| \frac{a(\La, \kappa)}{\sqrt \La}\right| <C_{\rm max}.$$
\el \proof Note that for a continuous function $f$, \eq{SF}
\frac{d}{d\La } \rrrr  f(r_1, r_2) = \int_\kappa^\La f(\La,r) {\rm
d}r+\int_\kappa^\La f(r,\La) {\rm d}r. \en In this proof,  $C$
denotes some sufficiently large constant and is not necessarily
the same number. \medskip\\
{\bf (1)} We have
$$ \frac{d}{d\La} b_1(\La)
=8\pi  \Y \rr\lk\frac{\La}{r+2}+\frac{r}{\La+2}\rk(1+X^2). $$
Since by Lemma \ref{hir1} (3) and (1),
\be \Y\rr\frac{\La}{r+2}&\leq& C \frac{\log \La}{\La}, \\
\Y\rr\frac{r}{\La+2} &\leq& C \frac{1}{\La},
\ee
 we have
\eq{Mame1}
\left|\frac{d}{d\La} b_1(\La)\right|\leq C \frac{\log \La}{\La}.\en
{\bf (2)} We see that by Lemma \ref{hir1} (2),
\be
\frac{d}{d\La} b_2(\La) &=&8 \pi \Y \lk \rr\rk^3 r\La  (\La^2+r^2+2\La rX)(1+X^2)\\
&\leq&  8\pi  \Y \lk\rr\rk^2 r \La  (1+X^2)\\
&\leq& 8\pi \La ^2 \Y \lk\rr\rk^2(1+X^2)\\
&\leq& C \frac{1}{\sqrt{\La }} .
\ee
Hence
\eq{103}
\left|
\frac{d}{d\La } b_2(\La )
\right| \leq C \frac{1}{\sqrt \La }.
\en
{\bf (3)} We have
$$\lz \frac{d}{d\La } b_3(\La )\rz
=16\pi  \lz
\Y X(X^2-1) \lk \rr\rk^2\lk \frac{\La ^2 r}{r+2}+\frac{r^2 \La }{\La +2}\rk\rz.$$
Since by Lemma \ref{hir1} (3) and (4),
\be
&& \left|
\Y X(X^2-1) \lk\rr\rk^2 \frac{\La ^2 r}{r+2}\right|
\leq C \frac{1}{\La },\\
& & \left|
\Y X(X^2-1) \lk\rr\rk^2 \frac{r^2\La }{\La +2}\right|
\leq C \frac{1}{\La },
\ee
we have
\eq{Mame2}
\left|\frac{d}{d\La } b_3(\La )\right|\leq
C \frac{1}{\La }.\en
{\bf (4)} It is trivial that
\eq{106}
\left| b_4(\La )\right| \leq C [\log \La ]^2.\en
{\bf (5)} We have
$$
\frac{d}{d\La } b_5(\La )
=
16\pi
\Y \rr (1-X^2) r\La \lkk\lk\frac{1}{\La +2}\rk^2+\lk\frac{1}{r+2}\rk^2 \rkk.$$
Since by Lemma \ref{hir1} (1),
\be
& &
\Y \rr (1-X^2) r \La  \lk\frac{1}{\La +2}\rk^2
\leq C \frac{1}{\La },\\
& &
\Y \rr (1-X^2) r\La   \lk\frac{1}{r+2}\rk^2
\leq  C\frac{1 }{\La },
\ee
we see that
\eq{107}
\left|
\frac{d}{d\La } b_5(\La )
\right|\leq  C \frac{1}{\La }.
\en
{\bf (6)} We have by Lemma \ref{hir1} (1)
$$\left|
\frac{d}{d\La }
b_6(\La ) \right|
=16\pi  \left|
\Y (X^2-1)X \rr \frac{r}{r+2}\frac{\La }{\La +2}\right|
\leq C \frac{1}{\La }.$$
Then we have
\eq{109}
\left|
\frac{d}{d\La }b_6(\La )
\right|\leq C \frac{1}{\La }.
\en
From \kak{Mame1}-\kak{109} it follows that
$$|b_j(\La )|\leq C [\log \La ]^2,\ \ \ j=1,4,$$
\eq{sa3}
|b_2(\La )|\leq C  \La ^\han,\ \ \
|b_j(\La )|\leq C \log \La , \ \ \ j=3,5,6.
\en
Then the lemma follows.
\qed

\subsection{Lower bounds}
\bl{hiro2}
There exists a positive constant $C_{\rm min}>0$ such that
$$C_{\rm min}  \leq \limR \frac{b_2(\La )}{\sqrt \La } .$$
\el
From this lemma and  \kak{sa3}, i.e.,
$$\limR  \frac{b_j(\La )}{\sqrt \La }=0,\ \ \ j=1,3,4,5,6,$$
the following corollary follows.
\bc{sa2}
It follows that
$$C_{\rm min} \leq \limR \frac{a_2(\La ,\kappa)}{\sqrt \La }.$$
\ec
{\it Proof of Lemma \ref{hiro2}:} We have
\eq{sa1}
\frac{d}{d\La } b_2(\La )=8\pi\Y (1+X^2) \lk\rr\rk^3 (r^2+2r\La X+\La ^2)r\La ,
\en
where, recall that
\be
\rho_\La (r,X)&=&
(r+\La X+1)^2+\Delta,\\
\Delta &= &\La ^2(1-X^2)+\La (1-X)-1.\ee
Note that $\Delta>0$ for $X\leq 0$ and a sufficiently large $\La $.
Since the integrand of \kak{sa1}
$$\trr=\lk\rr\rk^3(r^2+2r\La X+\La ^2)r\La $$ is positive,
it is enough to prove that
\eq{W1}
\limR\sqrt \La  \frac{d}{d\La }b_2(\La )=
\limR \sqrt \La  \YY
\trr (1+X^2)>0 .
\en
We simply set
$\rrr=\rho_\La (r, X)$.
Since
$$(r^2+2r\La X+\La ^2)r\La
=\La\lkk (r-2)\rrr+(4+4\La X-2\La )r+2(\La ^2+2\La )\rkk,$$
we have
\be
\II \trr
&=&
\La \II \frac{r-2}{\rrr^2}+\La (4+4\La X-2\La )\II\frac{r}
{\rrr^3}+2\La (\La ^2+2\La )\II\frac{1}{\rrr^3}\\
&=& \La \II\frac{r}{\rrr^2}+\La ^2(4X-2)\II\frac{r}{\rrr^3}+2\La^3
\II\frac{1}{\rrr^3}+t_1(\La ), \ee where
$$t_1(\La )=-2\La \II\frac{1}{\rrr^2}+4\La \II\frac{r}{\rrr^3}+2\La ^2\II\frac{1}{\rrr^3}.$$
Moreover
since
\be
& & \II\frac{r}{\rrr^2}=-\half\M{\frac{1}{\rrr}}-\II\frac{\La X+1}{\rrr^2},\\
& & \II\frac{r}{\rrr^3}=-\frac{1}{4}\M{\frac{1}{\rrr^2}}-\II\frac{\La X+1}{\rrr^3},
\ee
we have
$$\II \trr
=
-\La ^2 X \II \frac{1}{\rrr^2}+\La ^3(2-X(4X-2)) \II\frac{1}{\rrr^3}+t_1(\La )+t_2(\La ),$$
where
\be
t_2(\La ) &=& \La (-\half) \M{\frac{1}{\rrr}}-\La \II\frac{1}{\rrr^2}\\
& & +\La ^2(4X-2)(-\frac{1}{4})\M{\frac{1}{\rrr^2}}-\La ^2(4X-2)\II\frac{1}{\rrr^3}.
\ee
Note that
$$\int \frac{1}{(x^2+a^2)^{n+1}} {\rm d}x=
\frac{1}{2n a^2}\frac{x}{(x^2+a^2)^n} +
\frac{2n-1}{2n}\frac{1}{a^2}\int\frac{1}{(x^2+a^2)^n} {\rm d}x,\ \
n\geq 1.$$ Then \be & & \II\frac{1}{\rrr^2}=\M{\frac{r+\La
X+1}{2\Delta}\frac{1}{\rrr}}+
\frac{1}{2\Delta^{3/2}} \N,\\
& & \II\frac{1}{\rrr^3}=\M{\frac{r+\La X+1}{4\Delta}\frac{1}{\rrr^2}}+\frac{3}{8}
\M{\frac{r+\La X+1}{\Delta^2}\frac{1}{\rrr}}\\
& & \hspace{3cm}+\frac{3}{8}\frac{1}{\Delta^{5/2}}
\N.
\ee
Hence
we have
\be
& & \II \trr
=
-\La ^2 X \frac{1}{2\Delta^{3/2}}\lkk \N+\RN\rkk \\
& & +\frac{3}{8} \La ^3 2 (2X+1)(1-X)
\frac{1}{\Delta^{5/2}}\lkk \N+\RN\rkk\\
& & \hspace{3cm}
+t_1(\La )+t_2(\La )+t_3(\La ),
\ee
where
$$t_3(\La )=\La ^3 2 (2X+1)(1-X)
\M{\frac{r+\La X+1}{4\Delta}\frac{1}{\rrr^2}}.$$
It is proven  in Lemma \ref{W2} of Appendix B that
\eq{WW}
\limR \sqrt \La  \int_{-1+1/\La }^0 (1+X^2)( t_1(\La )+t_2(\La )+t_3(\La )) dX=0.
\en
From this it is enough to show that
$$\limR \sqrt \La  \La ^2 \int_{-1+1/\La }^0 dX (1+X^2) \frac{1}{\Delta^{3/2}} \RN
\times  $$
\eq{po2}
\times
\frac{1}{4} \lkk -2X+3(1-X)(2X+1)\frac{\La }{\Delta}\rkk\geq 0
\en
and that
there exists a  positive constant $\xi>0$ such
that
$$\limR \sqrt \La   \int_{-1+1/\La }^0 dX (1+X^2)
\lkk -\La ^2 X\frac{1}{2\Delta^{3/2}}\N\right.$$
$$\left. + \frac{3}{8} \La ^3 2 (2X+1)(1-X)
\frac{1}{\Delta^{5/2}}\N\rkk$$
$$=\limR \sqrt \La  \La ^2 \int_{-1+1/\La }^0 dX (1+X^2) \frac{1}{\Delta^{3/2}} \N  \times $$
\eq{po1}
\times
\frac{1}{4} \lkk -2X+3(1-X)(2X+1)\frac{\La }{\Delta}\rkk>\xi
\en

Changing variable $X$ to $-y$, we shall prove \kak{po1}, i.e.,
$$ \limR \sqrt \La  \La ^2 \YYY  (1+y^2) \frac{1}{\Delta^{3/2}} \NN \frac{1}{4}
a_\La (y) dy>\xi, $$
where
\be
& & a_\La (y)=2y+\frac{6}{\La }+b_\La (y),\\
& & b_\La (y)=\frac{3}{\La }\lk 1+\frac{2}{\La }\rk
\frac{y+\frac{2\La +3} {2\La +4}}{\lk y-\frac{1}{2\La
}\rk^2-\frac{(2\La +3)(2\La -1)}{4\La ^2}}. \ee The function
$b_\La (\cdot)$ satisfies the following properties: \bi
\item[(1)] $b''_\La (y)<0$ for $0\leq y\leq 1-1/\La $,
i.e., $b_\La (y)$ is concave for $0\leq y\leq 1-1/\La $,
\item[(2)] $\limR b_\La (1-1/\La )=-3/2$,
\item[(3)] $\limR b_\La (y)=0$ for $y\not=1$ and $\limR b_\La (1)=-3$.
\ei
By (1)--(3) we have
$$\inf_{0\leq y\leq 1-1/\La }b_\La (y)=\min\lkk b_\La (0), b_\La (1-1/\La )\rkk$$
and then for sufficiently large $\La $,
$$\inf_{0\leq y\leq 1-1/\La }b_\La (y)= b_\La (1-1/\La )>-\frac{7}{4}.$$
Hence
$$\inf_{15/16\leq y\leq 1-1/\La } a_\La (y)\geq \frac{15}{8}-\frac{7}{4}=\frac{1}{8}>0.$$
Moreover
\be
& &  \NN=\arctan\frac{(1-y)\La +1}{\sqrt \Delta}-\arctan\frac{\kappa-\La y+1}{\sqrt\Delta}>0,\\
& & \limR \NN=\arctan\frac{1-y}{\sqrt{1-y^2}}+\arctan\frac{y}{\sqrt{1-y^2}}>0
\ee
for $0\leq y\leq 1.$
Then
$$\delta=\inf_{\La >1} \inf_{0\leq y\leq 1} \NN>0$$
Then we have
\be
& & \limR \sqrt \La  \La ^2 \YYY (1+y^2) \frac{1}{\Delta^{3/2}} \NN \frac{1}{4} a_\La (y)\\
& & \geq \limR  \sqrt \La  \La ^2 \YYYY (1+y^2) \frac{1}{\Delta^{3/2}} \NN \frac{1}{4} a_\La (y)\\
& & \geq  \frac{1}{8} \times  \frac{1}{4} \times \delta \times \limR \sqrt \La  \La ^2 \YYYY  \frac{1}{\Delta^{3/2}}.
\ee
Furthermore
\be
& & \limR \sqrt \La  \La ^2 \YYYY \frac{1}{\Delta^{3/2}}\\
& & =
\limR \frac{1}{\sqrt \La } \YYYY \frac{1}{\lkk (1-y^2)+\frac{1}{\La }(1+y)-\frac{1}{\La ^2}\rkk^{3/2}}\\
& & \geq
\limR \frac{1}{\sqrt \La } \YYYY \frac{1}{\lkk (1-y)+\frac{1}{\La } \rkk^{3/2}}\frac{1}{(1+y)^{3/2}}\\
& &
\geq \limR \frac{1}{\sqrt \La } \YYYY \frac{1}{\lkk (1-y)+\frac{1}{\La } \rkk^{3/2}}\frac{1}{2^{3/2}}\\
& & =\limR \frac{1}{\sqrt 2}\frac{1}{\sqrt \La }
\lk \frac{1}{\sqrt{2/\La }}-\frac{1}{\sqrt{1/16+1/\La }}\rk=\half.
\ee
Then we proved that
$$\limR \sqrt \La \frac{d}{d\La } b_2(\La )> 4\pi \times
\frac{1}{8}\times  \frac{1}{4} \times \half\times \delta =
\frac{\pi \delta}{16}>0.$$
Then \kak{po1} follows.
We shall show \kak{po2}.
Since the left-hand side of \kak{po2} is
\eq{po3}
 \limR \sqrt \La  \La ^2 \YYY  (1+y^2) \frac{1}{\Delta^{3/2}} \RNN \frac{1}{4}
a_\La (y) dy,
\en
it is enough to show that $\M{\cdots}$ in \kak{po3} is nonnegative.
We can directly see that
$$ \RNN = \frac{\sqrt \Delta K}{\lkk(\La +\La X+1)^2+\Delta\rkk\lkk(\kappa+\La X+1)^2+\Delta\rkk},$$
where, for $0\leq y\leq 1$,
$$K=(-2y^2+y+1)\La ^3+(1+4y)\La ^2-2 \La  $$
$$+\kappa((y^2-2)\La ^2+(-2y-2)\La +1) + \kappa^2((-y+1)\La +1).$$
Since $K>0$ for a sufficiently large $\La $,  \kak{po2} follows. \qed\noindent
{\it Proof of Theorem \ref{main}:} The theorem follows from Lemma \ref{H3} and Corollary \ref{sa2}.\qed
\begin{remark}
(1)
$a_2(\La,\kappa)/\sqrt\La$ converges to a nonnegative constant as $\La\rightarrow\infty$.\\
(2) By \kak{imamiya}, we can define $a_2(\La, 0)$
since $b_j(\La)$ with $\kappa=0$ are finite.
Moreover $a_2(\La, 0)$ also satisfies Theorem \ref{main}.
\end{remark}
\appendix
\noindent {\Large {\bf Appendix}}
\section{Proof of Lemma \ref{hir1}}
{\it Proof of Lemma \ref{hir1}}\\
By the definition of $\D$ it follows that for
sufficiently large $\La $,
$$\frac{1}{\Delta}\leq \frac{1}{\La },\ \ \ {\rm for}\ X\leq 0.$$
Let
$$\delta =\delta(k)=\frac{1}{\La ^k},\ \ \ 0< k\leq 2.$$
Then for sufficiently large $\La $,
$$\Delta\geq \La ^2(1-X^2)-1>0,\ \ \ {\rm for}\
-1+\delta(k) <X\leq 0,\ \ \ 0<k\leq2.$$
In particular we obtain
$$\frac{1}{\Delta}\leq \frac{a}{\La ^2}
\frac{1}{{1-X^2}},\ \ \ {\rm for}\ -1+\delta(k) \leq X\leq 0,\ \ \ 0<k\leq 2,$$
with some constant $a$ independent of $\La $.
In this proof $C$ denotes some sufficiently large constant
and it is not necessarily the same number.
We divide  $\int_{-1}^1\cdots {\rm d}X$ as
$$\int_{-1}^1\cx  =\int_0^1 +\int_{-1+\delta }^0 +\int_{-1}^{-1+\delta } .$$
{\bf (1)}
It is trivial that
$\displaystyle \lz \int_0^1 \cx \rz\leq \frac{C}{\La }.$
Note that
\be
\lz \int_0^\La \dr \rr\rz
&=&
\frac{1}{\sqrt\Delta}\lz \arctan\frac{\La +\La X+1}
{\sqrt\Delta}-\arctan\frac{\La X+1}{\sqrt\Delta}\rz
\\
&\leq& \pi \frac{1}{\sqrt\Delta}.
\ee
Let $\delta =\delta(1/2)=1/\sqrt \La $.
Hence we have
\be
\lz \int_{-1+\delta }^0\cx \rz &\leq& \frac{C}{\La }\arcsin(1-\delta ),\\
\lz \int_{-1}^{-1+\delta }\cx \rz &\leq& \frac{C}{\sqrt \La } \delta .
\ee
Thus (1) follows.\medskip\\
{\bf (2)} It is trivial that
$\displaystyle \lz \int_0^1 \cx \rz\leq \frac{C}{\La ^3}.$
Note that
\be
& & \lz \int_0^\La  \dr \frac{1}{\rho_\La (r,X)^2} \rz\\
& & =\frac{1}{2\Delta}\lz
\int_0^\La
\rr \dr  +\lk \frac{\La +\La X+1}{(\La +\La X+1)^2+\Delta}-\frac{\La X+1}
{(\La X+1)^2+\Delta}\rk\rz\\
& & \leq \lkk
\begin{array}{ll}
\frac{C}{\La ^2}\lk\frac{1}{\La (1-X^2)^{3/2}}+\frac{1}{\La }\rk,& -1+\delta \leq X\leq 0,\\
\frac{C}{\La }\lk\frac{1}{\sqrt \La }+\frac{1}{\La }\rk,& -1\leq X\leq -1+\delta .
\end{array}
\right.
\ee
Let $\delta =\delta(1)=1/\La $.
Hence we have
\be \lz \int_{-1+\delta }^0\cx \rz &\leq& \frac{C}{\La ^3}
\int_{-1+\delta }^0 {\rm d} X \lk
\frac{1}{(1-X^2)^{3/2}}+1\rk\leq \frac{C}{\La ^3}\lk\frac{1}{\sqrt\delta }+1 \rk\\
\lz \int_{-1}^{-1+\delta }\cx \rz &\leq& \frac{C}{\La }\lk\frac{1}
{\sqrt \La }+\frac{1}{\La }\rk\delta . \ee Then (2) follows.\medskip\\
{\bf (3)}
We see that
$$\rr\frac{1}{r+2}=\frac{l_1}{r+2}+\frac{l_2}{\rho_\La (r,X)},$$
where
$$l_1=\frac{1}{\La ^2}\frac{1}{(4X-1)/\La -1},\ \ \
l_2=\frac{1}{\La ^2}\frac{r+2\La X}{(4X-1)/\La -1}.$$
We have
\be
& & \lz \XX\int_0^\La  \dr \frac{l_1}{r+2}\rz\leq \frac{\log(\La +2)}{\La ^2}
\XX\frac{1}{(4X-1)/\La -1}\leq C\frac{\log \La }{\La ^2},\\
& &
\lz \XX\int_0^\La  \dr \frac{l_2}{\rho_\La (r,X)}\rz\leq \frac{\La }{\La ^2}
\XX\int_0^\La  \rr \frac{1+2X}{(4X-1)/\La -1}\leq \frac{C}{\La ^2}.
\ee
Hence (3) follows.\medskip\\
{\bf (4)} It is trivial that
$\displaystyle \lz \int_0^1 \cx \rz\leq \frac{C}{\La ^3}.$
Let $\delta =\delta(3/2)=1/\La ^{3/2}$.
From the proof of (2) it follows that
\be
\lz\int_{-1+\delta }^0\cx \rz &\leq& \frac{C}{\La ^3}
\int_{-1+\delta }^0 d{\rm X} \lkk
\frac{(1-X^2)}{(1-X^2)^{3/2}}+(1-X^2)\rkk\\
&\leq& \frac{C}{\La ^3}\lk \arcsin(1-\delta )+1\rk, \\
\lz\int_{-1}^{-1+\delta }\cx \rz&\leq& \frac{C}{\La }\lk\frac{1}
{\sqrt \La }+\frac{1}{\La }\rk\delta . \ee Hence (4) follows.
\qed
\section{Proof of  \kak{WW} }
\bl{W2}
We have
\eq{WWW}
\limR \sqrt \La  \int_{-1+1/\La }^0 (1+X^2)( t_1(\La )+t_2(\La )+t_3(\La )) dX=0,
\en
where
\begin{eqnarray*}
& & t_1(\La )=-2\La \II\frac{1}{\rrr^2}+4\La \II\frac{r}{\rrr^3}+2\La ^2\II\frac{1}{\rrr^3},\\
& & t_2(\La )= \La (-\half) \M{\frac{1}{\rrr}}-\La \II\frac{1}{\rrr^2},\\
& & t_3(\La )=\La ^3 2 (2X+1)(1-X)
\M{\frac{r+\La X+1}{4\Delta}\frac{1}{\rrr^2}}.
\end{eqnarray*}
\el \proof In this proof $C$ also denotes some sufficiently large
constant, which is not necessarily the same number. We have \be &
&
\sqrt \La  \La  \lz \YY \frac{1}{\rho^2} \rz\leq C \sqrt \La  \La  \frac{1}{\La ^{5/2}} = C\frac{1}{\La },\\
& & \sqrt \La  \La  \lz \YY \frac{r}{\rho^3} \rz\leq C \sqrt \La  \La ^2 \frac{1}{\La ^{7/2}}= C\frac{1}{\La },\\
& & \sqrt \La  \La ^2 \lz \YY \frac{1}{\rho^3} \rz\leq C \sqrt \La  \La ^2 \frac{1}{\La ^{7/2}}=
C\frac{1}{\La }.
\ee
Then
$$\limR \sqrt \La  \int_{-1+1/\La }^0 \dx (1+X^2) t_1(\La )=0$$
follows.
Next we shall show that
\eq{sha}
\limR \sqrt \La  \int_{-1+1/\La }^0 \dx (1+X^2) t_2(\La )=0.\en
Note that
$$\lz \M{\frac{1}{\rho}}\rz
=\lz \frac{1}{(\La +\La X+1)^2+\Delta}-\frac{1}{(\kappa+\La X+1)^2+\Delta}
\rz\leq \frac{2}{\Delta}\leq \frac{2}{\La ^2}\frac{1}{1-X^2}.$$
Then
$$\lz \int_{-1+1/\La }^0 \dx \M{\frac{1}{\rho}} \rz \leq C \frac{\log \La }{\La ^2}.$$
Similarly we can see that
$$\lz \int_{-1+1/\La }^0 \dx \M{\frac{1}{\rho^2}}\rz \leq C \frac{1}{\La ^3},$$
which implies that
\be
& & \sqrt \La  \La  \lz \int_{-1+1/\La }^0{\rm d}X \M{\frac{1}{\rho}} \rz\leq C \frac{\log \La }{\sqrt \La },\\
& & \sqrt \La  \La ^2 \lz \int_{-1+1/\La }^0{\rm d}X \M{\frac{1}{\rho^2}} \rz\leq C \frac{1}{\sqrt \La }.
\ee
Hence \kak{sha} follows.
Finally we shall show that
\eq{fi}
\limR \sqrt \La  \int_{-1+1/\La }^0 \dx (1+X^2) t_3(\La )=0.\en
We divide $\displaystyle \int_{-1+1/\La }^0 \dx $ as
$$\int_{-1+1/\La }^0\dx =\int_{-1+1/\La }^{-\han} \dx +\int_{-\han}^0\dx .$$
Since
$$\frac{1}{\Delta}\leq \frac{1}{\La ^2}\frac{1}{1-X^2}\leq \frac{1}{\La ^2}\frac{4}{3},
\ \ \ {\rm for}\ -\half\leq X\leq 0,$$
we see that
\eq{fi1}
\sqrt \La   \La ^3 \lz \int_{-\han}^0 \dx \M{\frac{r+\La X+1}{\Delta}\frac{1}{\rho^2}}\rz
\leq C\sqrt \La  \La ^3 \frac{\La }{\La ^2}\frac{1}{\La ^3}=C\frac{1}{\sqrt \La }.
\en
On the other hand
\be
& & \M{\frac{r+\La X+1}{\Delta}\frac{1}{\rho^2}}\\
& & =
\frac{\La +\La X+1}{\Delta}\frac{1}{\lkk(\La +\La X+1)^2+\Delta\rkk^2}-\frac{\kappa+\La X+1}{\Delta}
\frac{1}{\lkk(\kappa+\La X+1)^2+\Delta\rkk^2}.
\ee
Since
$$
 \lz \frac{\La +\La X+1}{\Delta}\rz\leq \frac{C}{\La },\ \ \
 \lz \frac{\kappa+\La X+1}{(\kappa+\La X+1)^2+\Delta} \rz \leq\frac{C}{\La },
$$
we have
$$\lz \int_{-1+1/\La }^{-\han} \dx \M{\frac{r+\La X+1}{\Delta}\frac{1}{\rho^2}}\rz \leq
C\frac{1}{\La } \int_{-1+1/\La }^{-\han}\frac{1}{\Delta^2}\leq
C\frac{1}{\La ^4}.$$ Then we obtain that \eq{fi2} \sqrt \La  \La
^3 \lz \int_{-1+1/\La }^{-\han} \dx \M{\frac{r+\La
X+1}{\Delta}\frac{1}{\rho^2}}\rz \leq C\frac{1}{\sqrt \La }. \en
Thus \kak{fi} follows from \kak{fi1} and \kak{fi2}. \qed\noindent
{\footnotesize {\bf Acknowledgment} {\footnotesize We thank K. R. Ito for 
useful comments and for a careful reading of the first manuscript. F. H. thanks a
kind hospitality of TU M\"unchen in 2003. 
F. H. also thanks  Grant-in-Aid 13740106
for Encouragement of Young Scientists and 
Grant-in-Aid for Science Reserch (C) 15540191 from MEXT 
for financial support.}

}

\begin{thebibliography}{99}



\bibitem{hisp2}
F.  Hiroshima and H.  Spohn,
Ground state degeneracy of the Pauli-Fierz model with   spin,
{\it Adv. Theor. Math. Phys.} {\bf 5} (2001), 1091--1104.


\bibitem{ne3}
E.  Nelson,   Interaction of nonrelativistic particles with a
quantized scalar field,   {\it J.   Math.   Phys.  }{\bf  5}  (1964),   1190--1197.


\bibitem{hisp1}
F.  Hiroshima and H.  Spohn,
Enhanced binding through  coupling to a quantum field,
{\it Ann. Henri  Poincar\'e} {\bf 2} (2001), 1159--1187.


\bibitem{sp}
H.  Spohn,   Dynamics of Charged Particles and Their Radiation
Field, Cambridge University Press, 2004.




\bibitem{hiit}
F. Hiroshima and K. R. Ito, Mass renormalization in nonrelativistic QED with spin 1/2, 
preprint, 2004. 



\bibitem{sp5}
H.  Spohn,
Effective mass of the polaron: A functional integral approach,
{\it Ann.   Phys.  } {\bf 175}  (1987),   278--318.





\bibitem{lilo}
E. Lieb and M.   Loss,   Self-energy of electrons in
non-perturbative QED, preprint, 1999.



\bibitem{lilo2}
E.   Lieb and M.   Loss, A bound on binding energies and mass
renormalization in models of quantum electrodynamics, {\it J.
Stat. Phys.} {\bf 108}, 1057--1069 (2002).



\bibitem{ha1}
C. Hainzl, Increase of the binding energy of an electron by
coupling to a photon field, math-ph/0204052, preprint, 2002.

\bibitem{hasi}
C.  Hainzl and R.  Seiringer,
Mass Renormalization and Energy Level Shift in Non-Relativistic QED,
math-ph/0205044, preprint, 2002.



\bibitem{fr2}
J.  Fr\"ohlich,   Existence of dressed one electron states in a
class of persistent models, {\it Fortschritte der Physik} {\bf 22}
(1974),   159--198.








\bibitem{ch}
T. Chen, Operator-theoretic infrared renormalization and
construction of dressed 1-particle states in non-relativistic QED,
mp-arc 01-301, preprint, 2001.




\bibitem{rs4}
M.  Reed and B.  Simon,   {\it Methods of Modern Mathematical Physics IV},
Academic Press,   1978.



\end{thebibliography}
\end{document}